\documentclass[10pt,a4paper,twoside,twocolumn,american,prd,nofootinbib,superscriptaddress]{revtex4-1}
\usepackage{lmodern}

\usepackage[T1]{fontenc}
\usepackage[utf8]{inputenc}
\usepackage{graphicx}
\usepackage{color}
\usepackage{babel}
\usepackage{booktabs}
\usepackage{amsmath}
\usepackage{amssymb}
\usepackage{mathtools}
\usepackage[unicode=true,pdfusetitle,
 bookmarks=true,bookmarksnumbered=false,bookmarksopen=false,
 breaklinks=false,pdfborder={0 0 0},backref=false,colorlinks=true]
 {hyperref}
\hypersetup{
 citecolor=blue,filecolor=blue,linkcolor=blue,urlcolor=blue}

\makeatletter  % access to @ macros

 \@ifundefined{textcolor}{}
 {%
   \definecolor{BLACK}{gray}{0}
   \definecolor{WHITE}{gray}{1}
   \definecolor{RED}{rgb}{1,0,0}
   \definecolor{GREEN}{rgb}{0,1,0}
   \definecolor{BLUE}{rgb}{0,0,1}
   \definecolor{CYAN}{cmyk}{1,0,0,0}
   \definecolor{MAGENTA}{cmyk}{0,1,0,0}
   \definecolor{YELLOW}{cmyk}{0,0,1,0}
 }

\makeatother % end access to @ macros

\begin{document}

\title{Ray tracing the integrated Sachs-Wolfe effect through the light cones of the Dark Energy Universe Simulation -- Full Universe Runs}

\author{Julian Adamek}
\email{julian.adamek@qmul.ac.uk}
\affiliation{Astronomy Unit, School of Physics \& Astronomy, Queen Mary University of London, 327 Mile End Road, London E1~4NS, UK}
\affiliation{LUTH, UMR 8102 CNRS, Observatoire de Paris, PSL Research University, Universit\'e Paris Diderot, 5 place Jules Janssen, 92195 Meudon, France}
\author{Yann Rasera}
\affiliation{LUTH, UMR 8102 CNRS, Observatoire de Paris, PSL Research University, Universit\'e Paris Diderot, 5 place Jules Janssen, 92195 Meudon, France}
\author{Pier Stefano Corasaniti}
\affiliation{LUTH, UMR 8102 CNRS, Observatoire de Paris, PSL Research University, Universit\'e Paris Diderot, 5 place Jules Janssen, 92195 Meudon, France}
\affiliation{Sorbonne Universit\'e, CNRS, UMR 7095, Institut d'Astrophysique de Paris, 98 bis bd Arago, 75014 Paris, France}
\author{Jean-Michel Alimi}
\affiliation{LUTH, UMR 8102 CNRS, Observatoire de Paris, PSL Research University, Universit\'e Paris Diderot, 5 place Jules Janssen, 92195 Meudon, France}
\affiliation{Sorbonne Universit\'e, CNRS, UMR 7095, Institut d'Astrophysique de Paris, 98 bis bd Arago, 75014 Paris, France}
\date{\today}

\begin{abstract}
The late integrated Sachs-Wolfe (ISW) effect correlates the Cosmic Microwave Background (CMB) temperature anisotropies with foreground cosmic large-scale structures. As the correlation depends crucially on the growth history in the era of dark energy, it is a key observational probe for constraining the cosmological model. Here we present a detailed study based on full-sky and deep light cones generated from very large volume numerical N-body simulations, which allow us to avoid the use of standard replica techniques, while capturing the entirety of the late ISW effect on the large scales. We post-process the light cones using an accurate ray-tracing method and construct full-sky maps of the ISW temperature anisotropy for three different dark energy models. We quantify in detail the extent to which the ISW effect can be used to discriminate between different dark energy scenarios when cross-correlated with the matter distribution or the CMB lensing potential. We also investigate the onset of non-linearities, the so-called Rees-Sciama effect which provides a complementary probe of the dark sector. We find the signal of the lensing-lensing and ISW-lensing correlation of the three dark energy models to be consistent with measurements from the {\it Planck} satellite. Future surveys of the large-scale structures may provide cross-correlation measurements that are sufficiently precise to distinguish the signal of these models. Our methodology is very general and can be applied to any dark energy or modified gravity scenario as long as the metric seen by photons can still be characterized by a Weyl potential. \\
\end{abstract}
\maketitle

\section{Introduction}\label{sec:intro}
In a flat Friedmann-Lema\^{i}tre-Robertson-Walker (FLRW) universe, the presence of dark energy (DE) generates a distinct imprint on the Cosmic Microwave Background radiation through the late integrated Sachs-Wolfe effect \cite{SachsWolfe1967}. This effect originates in
the decay of the gravitational potentials associated with large-scale structure whose growth rate is altered by the increasingly fast expansion driven by the DE component. CMB photons traveling through the structures
gain a small energy variation that generates temperature anisotropies at large angular scales. This is not the case in a matter dominated universe where the cosmic expansion exactly compensates the growth rate of structures, rendering the gravitational potentials constant in time.
Because of this, the detection of the late ISW effect in a flat universe is a direct probe of DE. However, this signal needs to be disentangled from that of other effects contributing to the CMB temperature anisotropies. For instance, one can use the cross-correlation with tracers of the distribution of large-scale structure (LSS) \cite{CrittendenTurok1996}. 

The advent of CMB satellite experiments mapping the full sky distribution of temperature anisotropies such as the {\it Wilkinson Microwave Anisotropy Probe} (WMAP, \cite{WMAP}) and {\it Planck} \cite{Planck} have made possible the realization of cross-correlation analyses of the CMB temperature anisotropy maps with the large-scale distribution of structures from galaxy surveys. These studies have resulted in numerous detections of the late ISW signal \cite{ISWdetections,PlanckISW} and provided cosmological parameter constraints complementary to those inferred from other cosmic probes (see e.g. \cite{ISWanalyses}). 

The next generation of galaxy surveys such as LSST\footnote{www.lsst.org}, \textit{Euclid}\footnote{www.euclid-ec.org} and SKA\footnote{www.skatelescope.org} will map the distribution of cosmic structures over volumes of the universe much larger than those currently probed. These will allow for more precise measurements of the ISW and to test the properties of DE \cite{Pogosian2005,Douspis2008,Ballardini2017,Stolzner2018}. These programs will measure the clustering of galaxies over an unprecedented range of scales, covering large linear modes to small ones, where the non-linear dynamics of the gravitational collapse of matter induces time-variations of the gravitational potentials. Such non-linearities produce temperature anisotropies at small scales through the so called Rees-Sciama (RS) effect \cite{ReesSciama1968}. Due to the complexity of the non-linear regime of matter clustering, cosmological model predictions have to rely on N-body simulations. However, this is challenging for studies of the ISW-RS 
effect, which by spanning from large to small non-linear scales require large-volume, high-resolution simulations. 

In the past, several works in the literature have investigated the ISW-RS effect using N-body simulations. These studies differ not only in the characteristics of the N-body simulations used, but also in the adopted ray-tracing methods which are necessary to compute the temperature anisotropies by means of 
a numerical integration along the photon path \cite{Puchades2006,Cai2009,Smith2009,Cai2010,Watson2014,Carbone2016}.
For instance, \cite{Puchades2006} used high-resolution simulations of a relatively small volume which allowed them to study the RS effect only. In contrast \cite{Cai2010} used a simulation of similar resolution with a volume $64$ times larger, which allowed them to investigate both the ISW signal on the large scales and the RS contribution at small ones. Both these studies have determined the variation of the gravitational potential along the photon path from multiple patched copies of the simulated volumes. This methodology has the advantage that the variables of interests are continuous at the replica's boundaries due to periodic boundary conditions of the simulations. On the other hand, such replica induce spurious effects since a photon propagating along the direction parallel to any of the principal axes of the simulation box will encounter the same structure several times. Hence, in order to reduce such artifacts, the computation must be limited to oblique photon trajectories crossing the box replica 
at different points with a maximum radial depth set by the simulation box-length. This suppresses the contribution of such spurious effects, but it does not solve the problem entirely.
A different approach consists instead in patching repeated boxes that have been randomized by applying a random translation and rotation to the box coordinates. This avoids the repetition of structures along the photon path, but on the other hand it introduces artifacts due to the discontinuity at the replica boundaries. To limit such effects, \cite{Carbone2016} have used replica in which boxes contributing to the same redshift shell have undergone the same random translation and rotation (see \cite{Carbone2008}). Hence, each redshift shell around the observer has a different randomization. This reduces the effect of discontinuities at the boundaries of replica, while eliminating any preferred direction in the simulated sky-maps. Still, this approach does not account for physical correlations among structures residing in nearby redshift shells. Moreover, in all these works the integration along the photon trajectory is carried out interpolating data from snapshots of the simulation box at given redshift outputs. 
However, the number of redshift outputs of a simulation remains well below the actual time resolution of the simulations.

Here, we describe a comprehensive methodology which addresses all these limitations. To illustrate the strength of this approach we confront the ISW signal from numerical simulations to the prediction of the linear theory. We also investigate the imprint of non-linearities on the CMB-LSS correlation and its cosmological dependence on the DE model.
To this end we use the full-sky light-cone data from the Dark Energy Universe Simulation -- Full Universe Runs (DEUS-FUR) \cite{Alimi2012,Reverdy2015} in combination with a sophisticated ray-tracing technique that solves the photon geodesic equations along the photon trajectory. Our approach takes advantage of the fact that the DEUS-FUR simulations cover the volume of the entire observable universe allowing to generate full-sky light-cone data without the need of recurring to any replica method. Moreover, the light-cone data have been generated during the DEUS-FUR runtime which preserves the full time resolution of the simulations in the computation of the photon trajectories. Finally, the ray-tracing method allows us to integrate the geodesic equations along the perturbed photon path, rather than the usually adopted Born approximation along the unperturbed path. In the present article we focus mainly on large scales, nevertheless our methodology is very general and can be applied to smaller scales where non-linearities enhance the differences between predictions of different cosmological models.

The paper is organized as follows. In Section~\ref{sec:isw} we briefly review the equations underlying the ISW effect. In Section~\ref{sec:methodology} we will describe the numerical simulation datasets and the ray-tracing method. In Section~\ref{sec:results} we will present the results, while in Section~\ref{sec:conclusions} we will discuss our conclusions.

\begin{figure*}[tb]
\begin{minipage}{\textwidth}
 \centerline{\includegraphics[width=\columnwidth]{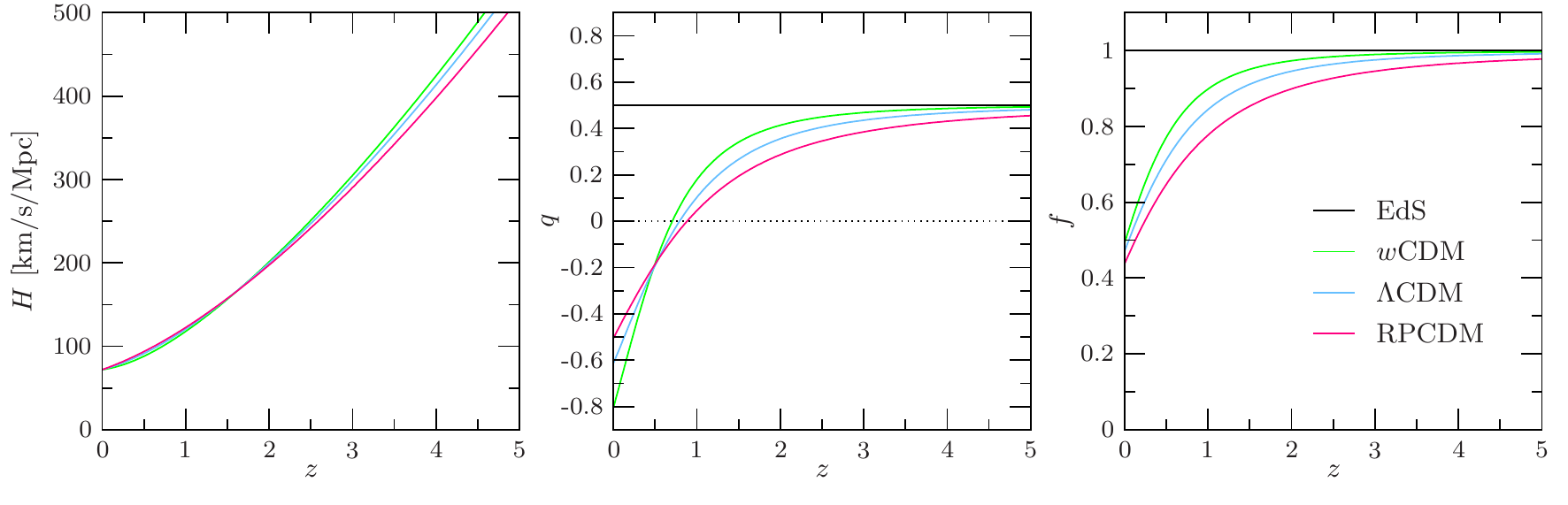}}
\end{minipage}
\caption{\label{fig:cosmo} \small  Hubble function $H$, deceleration parameter $q$ and linear growth rate $f$ as a function of redshift $z$ for the $\Lambda$CDM (blue solid lines), RPCDM (magenta solid lines) and $w$CDM (green solid lines) models, respectively. As reference for the latter two we plot as horizontal black lines the constant values of the Einstein--de~Sitter model with $\Omega_m=1$  ($q_{\rm EdS}=0.5$, $f_{\rm EdS}=1$). }
\end{figure*}
 
\section{ISW Effect}\label{sec:isw}
In order to gain an intuitive understanding of the various contributions to the energy of photons traversing structures in an expanding universe, let us consider a linearly perturbed FLRW metric in longitudinal gauge (given here in natural units),
\begin{equation}
 ds^2 = a^2(\tau) \left[-(1 + 2 \psi) d\tau^2 + (1 - 2 \psi) \delta_{ij} dx^i dx^j\right]\, ,
\label{metric}
\end{equation}
where $\tau$ is the conformal time, $a(\tau)$ is the scale factor and $\psi$ is the Newtonian potential (we neglect anisotropic stress, vector and tensor perturbations). In this coordinate system the energy of a photon $k^0$ evolves as
\begin{equation}
 \frac{dk^0}{d\tau} + 2 \mathcal{H} k^0 + 2 \frac{\partial\psi}{\partial x^i} n^i k^0 = 0\, ,
\end{equation}
where $\mathcal{H} = d\ln a / d\tau$ and $n^i$ is the unit vector pointing into the direction in which the photon is traveling. A first-order integral of this equation yields the following perturbative expression for the observed photon redshift:
\begin{widetext}
\vspace{-11pt}
\begin{equation}
 1 + z = \underbrace{\frac{a_\mathrm{obs}}{a_\mathrm{src}}}_{\mathclap{\mathrm{background}}} \biggl(1 + \underbrace{\mathbf{n}\!\cdot\!\mathbf{v}_\mathrm{obs} \underset{~}{-} \mathbf{n}\!\cdot\!\mathbf{v}_\mathrm{src}}_{\mathclap{\mathrm{Doppler}}} + \underbrace{\psi_\mathrm{obs} \underset{~}{-} \psi_\mathrm{src}}_{\mathclap{\mathrm{time~dilation}}} - \underbrace{2 \int\limits_\mathrm{src}^\mathrm{obs} \frac{\partial \psi}{\partial \tau} d\chi}_{\mathclap{\mathrm{ISW~effect}}}\biggr),
 \label{ISWEQ}
\end{equation}
\vspace{-10pt}
\end{widetext}
where $\mathbf{v}$ is the peculiar velocity, $d\chi$ is the conformal distance element (not the line element which is always null for a photon), the subscript `$\mathrm{src}$' refers to the source and the subscript `$\mathrm{obs}$' refers to the observer.

The measured energy of a CMB photon is an observable and hence does not depend on the coordinate system used to describe it. However, the advantage of using longitudinal gauge coordinates is that the terms appearing in the above expression have a clear interpretation. We may notice that apart from the factor related to the overall expansion of the background there are two types of contributions that are local at the source and observer, namely the Doppler shift due to peculiar motion and the gravitational redshift due to time dilation. The last term, which is an integral along the photon path, is precisely what we call the ISW-RS effect. Its origin is also quite intuitive: a photon will receive a net blueshift if it travels through a decaying ($\partial \psi / \partial \tau > 0$) potential well, since it gains more energy on the infall than it loses climbing out of the increasingly shallow 
well.

It is worth remarking that N-body simulations are usually not realized in the longitudinal gauge\footnote{At leading order the correct gauge for interpreting Newtonian simulations is presented in \cite{Fidler2015}.}. Hence, in numerical simulation analyses the full expression for the observed redshift should be properly gauge-transformed, which introduces some new local terms that can be of the same order as the gravitational redshift (see \cite{Adamek:2017kir}). However, the integrated contribution we are interested in here is readily gauge invariant, i.e.\ up to second-order corrections the integral can be taken directly in the coordinates of the Newtonian simulation.

\section{Methodology}\label{sec:methodology}
We use numerical data from the DEUS-FUR project taking advantage of the available full-sky and deep light cones (with no replica) for several observers and for three cosmological models.

\subsection{Cosmological Models}
\label{subsec:cosmomodels}
The cosmological models of the DEUS-FUR simulations consist of a flat $\Lambda$CDM model, a quintessence model with Ratra-Peebles potential (RPCDM, \cite{Ratra1988}) and a phantom dark energy model with constant equation of state $w<-1$ ($w$CDM, \cite{Caldwell2003}). The cosmological model parameters have been calibrated to fit the CMB anisotropy power spectra from the WMAP-7yr data \cite{Spergel2007} and the luminosity distance measurements from supernova Ia standard candles \cite{Kowalski2008}. These models are known as \emph{realistic} models \cite{Alimi2010}. In particular, the values of cosmic matter density $\Omega_m$ and the normalization of the root-mean-square fluctuations $\sigma_8$ of the non-standard dark energy models have been chosen within the $68\%$ confidence region along the degeneracy line of the $\Omega_m-w$ and $\sigma_8-w$ planes respectively, such as to be statistically indistinguishable from the $\Lambda$CDM best-fit model at $1\sigma$ (see Fig.\ 1 in \cite{Bouillot2015}). A summary of the cosmological model parameter values and the simulation characteristics is reported in Table \ref{DEUSSALL1}.

\setcounter{table}{0}
\begin{table}
\begin{center}
\begin{tabular}{@{\quad}c@{\quad}c@{\quad}c@{\quad}c@{\quad}}
\hline
Parameters & RPCDM & $\Lambda$CDM & $w$CDM \\
\hline
\hline
$\Omega_m$  & 0.23& 0.2573 & 0.275 \\
$\Omega_b h^2$  & 0.02273& 0.02258 & 0.02258 \\
$\sigma_8$& 0.66 & 0.8  & 0.852 \\
$w_0$& -0.87 & -1  & -1.2 \\
$w_a$ & 0.08 & 0 & 0 \\
\hline
$z_\mathrm{ini}$  & 94 & 106 & 107 \\
$m_p$ [M$_\odot$/$h$]  & $1.08 \times 10^{12}$ & $1.20 \times 10^{12}$  & $1.29 \times 10^{12}$ \\
$\Delta x$ [kpc/$h$] & 40 & 40 & 40 \\
\hline
\end{tabular}
\caption{Cosmological parameter values of the DEUS-FUR simulated cosmologies. For all models the scalar spectral index is set to $n_s=0.963$ and the Hubble parameter $h=0.72$. For information we also report the values of a linear equation of state parametrization $w(a) = w_0 + w_a(1- a)$ for the different models (though in the RPCDM case we have used the exact equation of state obtained by numerically solving the Klein-Gordon equation). In the bottom table we list the values of the initial redshift of the simulations $z_\mathrm{ini}$, the particle mass $m_p$ and the comoving spatial resolution  $\Delta x$. For all three simulations the box-length is L$_{\textrm {box}}=21000$~Mpc/$h$ and the number of N-body particles is $8192^3$.}

\label{DEUSSALL1}
\end{center}
\end{table}

The RPCDM and $w$CDM models are characterized by a background expansion and the linear growth of density fluctuations which bracket those of the $\Lambda$CDM. We can see this in Fig.~\ref{fig:cosmo}, where we plot the redshift evolution of the Hubble function (left panel), the deceleration parameter (center panel) and the linear growth rate (right panel) for the $\Lambda$CDM, RPCDM and $w$CDM models, respectively. First, we can see that the accelerated expansion starts earlier in RPCDM than $\Lambda$CDM (i.e. $z_{\rm RPCDM}^{\rm acc}>z_{\Lambda{\rm CDM}}^{\rm acc}$), while it starts later in $w$CDM. Also notice that during the preceding phase, the cosmic expansion in RPCDM is less decelerated than in $\Lambda$CDM (i.e. $q_{\rm RPCDM}<q_{\Lambda{\rm CDM}}$), while is more decelerated in $w$CDM. As we may notice from the evolution of the linear growth rate, this implies that structures will grow more efficiently in $w$CDM compared to the $\Lambda$CDM case (i.e. $f_{w{\rm CDM}}>f_{\Lambda{\rm CDM}}$) and less efficiently in RPCDM during the 
decelerated phase. These differences are also present during the subsequent accelerated phase of expansion, although exhibiting a different slope at low redshift due to the different rate of cosmic acceleration specific to each model. 

Given the fact that the variation of the gravitational potentials is proportional to the redshift variation of the growth rate of matter density fluctuations, the trends shown in Fig.~\ref{fig:cosmo} entirely characterize the cosmological dependence of ISW signal. While this will be discussed in detail in Section~\ref{sec:results}, for the time being it is informative to present an estimate of the ISW signal for the DEUS-FUR cosmologies assuming the linear theory. In particular, a simple expression of the ISW power spectrum can be obtained through the Limber approximation:
  \begin{equation}
    C^{TT}_\ell=\int_\mathrm{src}^\mathrm{obs} d\chi \frac{q_1(\chi) q_2(\chi)}{\chi^2} P_{\delta \delta}\left(\frac{\ell}{\chi},\chi\right), \label{eq:limber}
  \end{equation}
where $P_{\delta \delta}$ is the linear power spectrum and the weight functions
$q_i(\chi)$ are given by
\begin{equation}
  q_1(\chi)=q_2(\chi)=\frac{3 H_0^2 \Omega_m \chi^2 H}{\ell^2} (f-1),
\end{equation}
with $f$ being the linear growth rate. Eq.~(\ref{eq:limber}) provides an explicit relation between the linear growth rate and the ISW power spectrum.

\begin{figure}[tb]
\centerline{\includegraphics[width=0.89\columnwidth]{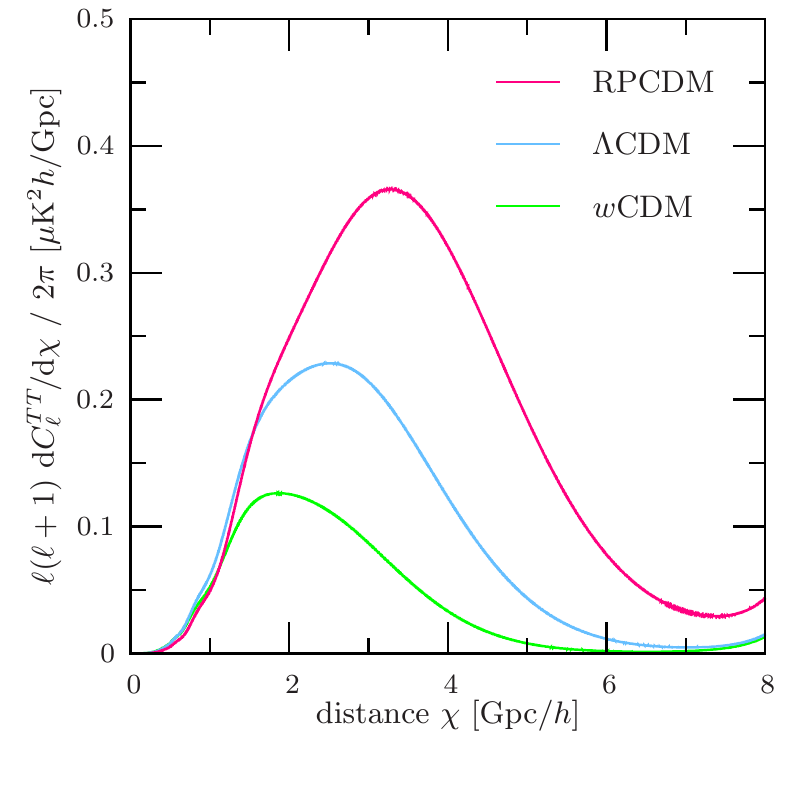}}
\caption{\label{fig:integrand} \small Contributions to the late ISW angular power spectrum at $\ell=100$ per unit of comoving distance. The computation assumes linear evolution as well as Limber approximation. The blue line shows the $\Lambda$CDM model contributions while red and green lines show the RPCDM and $w$CDM contributions. These contributions are directly related to the differences of linear growth rates with respect to EdS cosmology as shown Fig.~\ref{fig:cosmo}.}
\end{figure}

Using Eq.~(\ref{eq:limber}), we plot in Fig.~\ref{fig:integrand} the increment of
the linear ISW power spectrum per unit of comoving distance. 
We can see that this quantity varies in a cosmological model dependent way over very
large distances
and reaches a maximum around 2~Gpc/$h$ to about 4~Gpc/$h$. This implies that
the light cones built using N-body numerical simulations and which are used to estimate the ISW effect must be deep enough to include the full time variation of the integrand $\partial \psi / \partial \tau$. In particular, light cones must allow to integrate the time variation of the gravitational potential up to comoving distances of more than 7 to 8~Gpc/$h$.
As we will explain in the following subsections, the light cones generated from the DEUS-FUR simulations have optimal characteristics to perform such an integration.

In what follows, except otherwise stated, we will not rely on the Limber approximation for analytical calculations but rather on the full Bessel integrals as computed by the Boltzmann code CLASS \cite{Lesgourgues2011}.

\subsection{Simulations}
DEUS-FUR comprises three N-body simulations of a ($21$\, Gpc/$h$)$^3$ volume with $8192^3$ particles of a flat $\Lambda$CDM model and two DE scenarios with different expansion histories (see Table \ref{DEUSSALL1}). The simulations have been run using the application AMADEUS -- `A Multi-purpose Application for Dark Energy Universe Simulation' expressively developed for the realization of the DEUS-FUR project \cite{Alimi2012,Reverdy2015}. This includes the code generating Gaussian initial conditions for which we use an optimized version of MPGRAFIC \cite{Prunet2008}, the N-body solver for which we use a specifically modified version of the RAMSES code \cite{Teyssier2002} such as to run on a very large number of cores ($\approx 80000$) and a parallel friends-of-friends halo finder pFoF as described in \cite{Roy2014}. RAMSES solves the Vlasov-Poisson equations using an Adaptive Mesh Refinement (AMR) particle-mesh method with the Poisson equation solved with a multi-grid technique \cite{Guillet2011}. We refer the reader to
\cite{Alimi2012} and \cite{Reverdy2015} for a detailed description of the algorithms and optimization schemes adopted in the realization of the DEUS-FUR project.

At coarse level the grid of the DEUS-FUR simulations contains $8192^3$ cells, these are allowed to be refined six times reaching a formal spatial resolution of $\Delta{x}=40\,$ kpc/$h$, while the particle mass resolutions is $m_p\approx 10^{12}\,M_{\odot}/h$ for the different models. Although these are the first simulations covering the volume of the full observable Universe at such a resolution, this is still relatively low compared to that of other ISW studies in the literature, which have used smaller volumes \cite{Cai2010,Carbone2016}. Hence, our study will focus more on the large scales, where the DEUS-FUR simulations provide cosmic-variance limited errors \cite{Rasera2014}. 
We now discuss the characteristics of the light-cone data and the ray-tracing technique.

\subsection{DEUS-FUR Light Cones}\label{subsec:lightcone}

DEUS-FUR light cones were built on-the-fly using an onion-shell approach \cite{Fosalba2008,Teyssier2009}. At each coarse time step the shell at the appropriate conformal distance from the observer is recorded (thus moving successively closer and closer to the observer as the simulation advances in time).
The time steps multiplied by the speed of light are larger than the resolution of the spatial mesh, consequently 
each shell has a certain thickness
of many coarse cells. The total number of shells is of order $\sim 400-450$ depending on the cosmology. For each of the three cosmologies, five full-sky light cones (fraction of the sky $f_\mathrm{sky}=1$) were stored up to maximum redshift $z_\mathrm{max}=30$.
Each of the light cones corresponds to a specific space-time location of the observer given by the position vector ($x^1_\mathrm{obs}/L_\mathrm{box}$, $x^2_\mathrm{obs}/L_\mathrm{box}$, $x^3_\mathrm{obs}/L_\mathrm{box}$, $z_\mathrm{obs}$) where $x^i_\mathrm{obs}$ are the observer Cartesian coordinates in the simulation box going from $0$ to $L_\mathrm{box}$, $L_\mathrm{box}$ is the box length and $z_\mathrm{obs}$ is the redshift of the observer. The locations were chosen as  (0.5,0.5,0.5,0.), (0.5,0.1,0.1,0.), (0.1,0.5,0.9,0.), (0.5,0.5,0.5,0.5) and (0.5,0.5,0.5,1.2) so as to maximize the distance between the observers at $z=0$.
In this article we will focus on the two first light cones which do not overlap up to a redshift $z \sim 5.6$. For each light cone, two kinds of data are stored: particles and AMR cells. Particle light cones contain particle properties (position, velocity and redshift) while gravity light cones contain mesh properties (position, gravitational field, potential, density, AMR cell's child index). We use the particle light cones to compute maps of the comoving dark matter density on the unperturbed light cone. In contrast, given that our line element Eq.~({\ref{metric}}) is fully specified by the potential, we perform the ray tracing using the gravity light cones, which allow us to compute ISW-RS maps as well as weak-lensing maps.

\begin{figure*}[tb]
\begin{minipage}{\textwidth}
\begin{tabular}{cc}
  \includegraphics[width=0.48\textwidth]{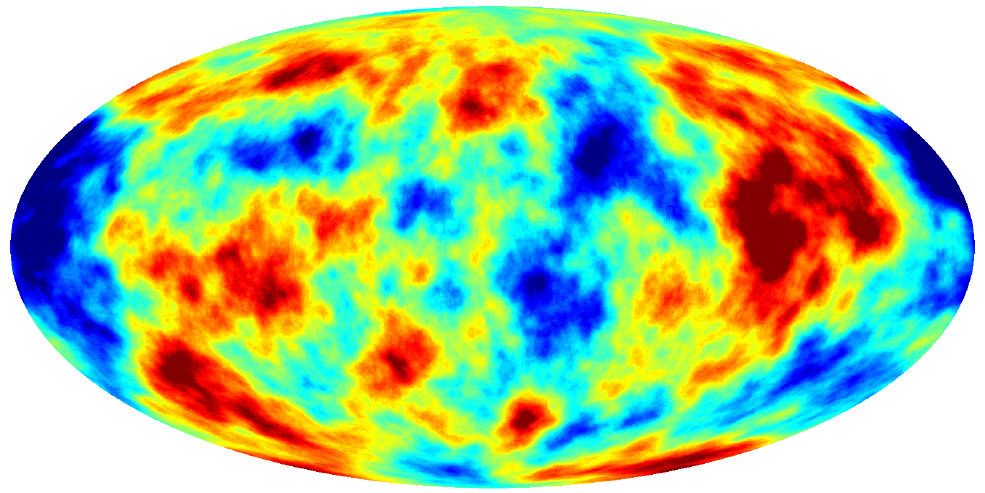} & \includegraphics[width=0.48\textwidth]{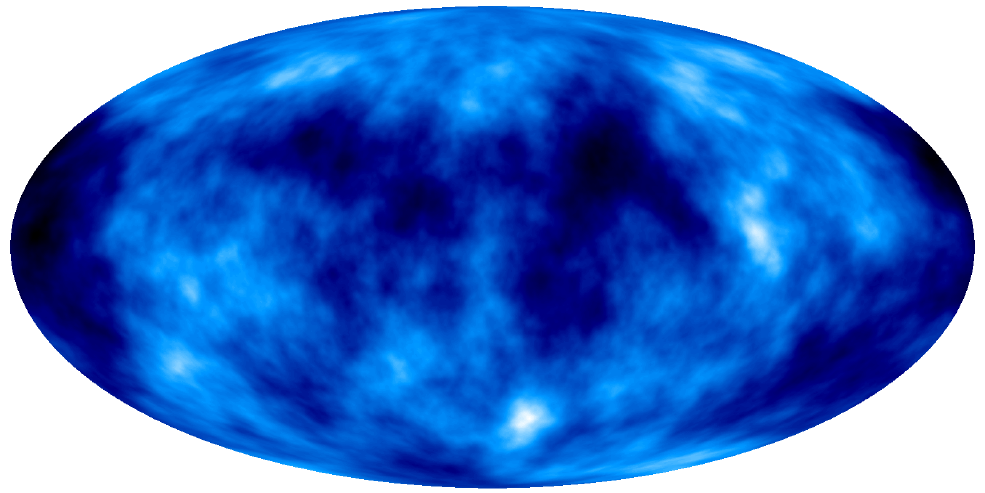} \\
 \includegraphics[width=0.48\textwidth]{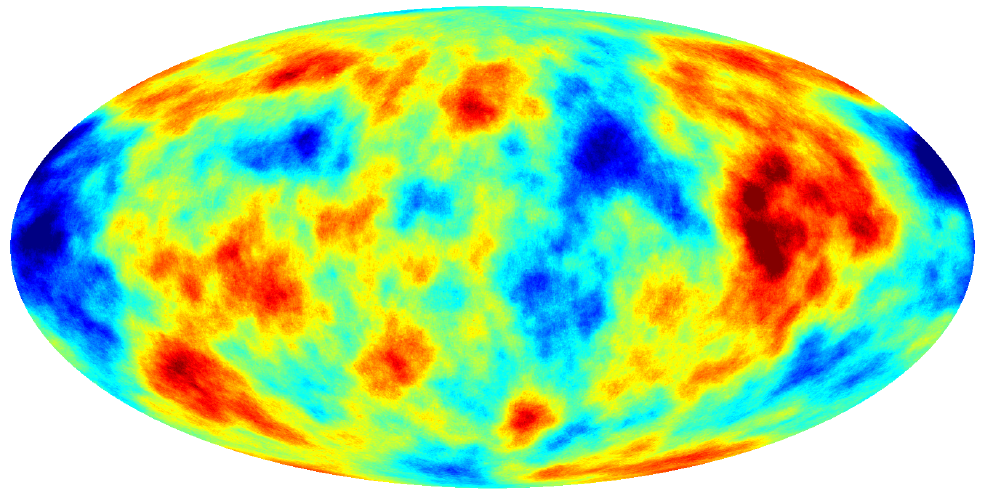} & \includegraphics[width=0.48\textwidth]{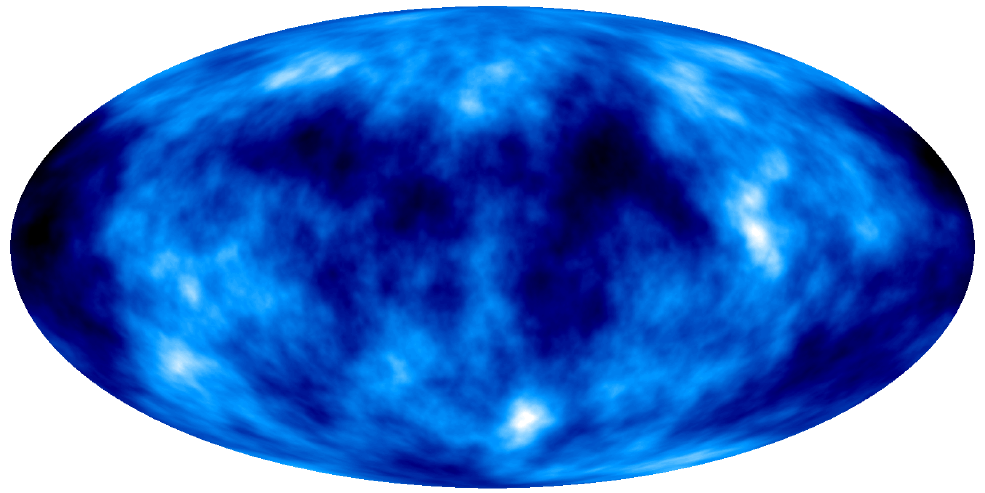} \\
 \includegraphics[width=0.48\textwidth]{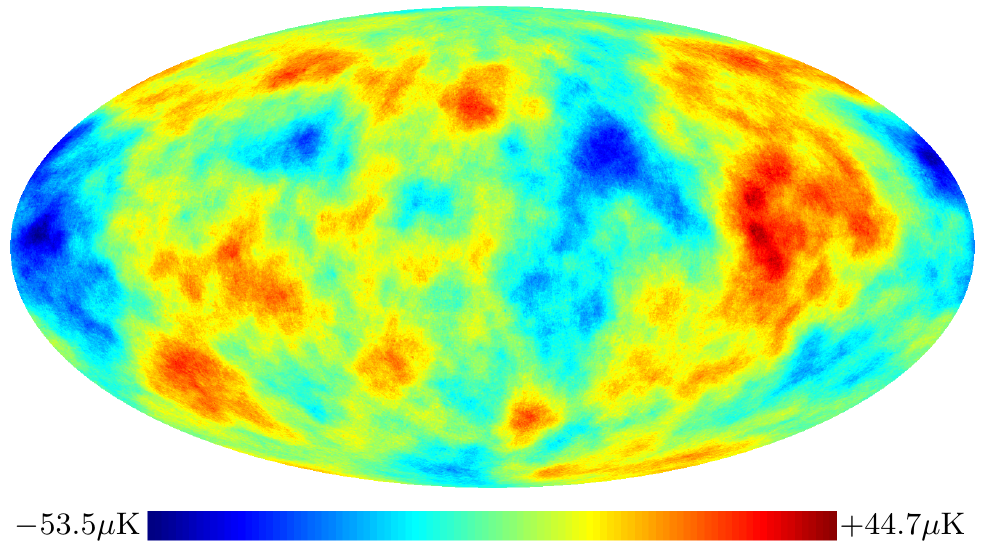} & \includegraphics[width=0.48\textwidth]{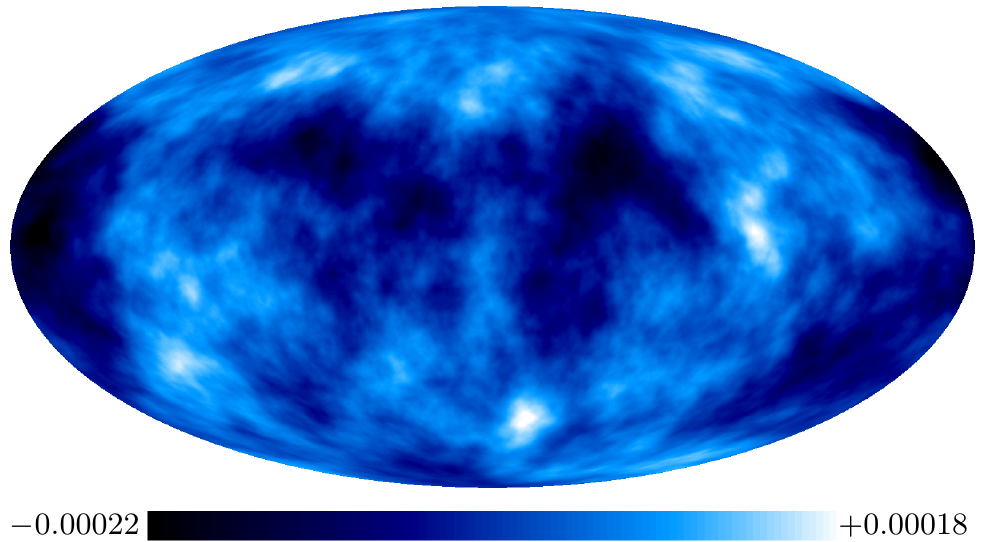}
\end{tabular}
\end{minipage}
\caption{\label{fig:maps} \small Full-sky maps of the ISW temperature anisotropy (left panels) and lensing potential (right panels) 
extracted from the DEUS FUR for (from top to bottom) RPCDM, $\Lambda$CDM, and $w$CDM models for one observer included in the simulation box. As usual, the temperature dipole is not shown since its observation is difficult due to the large kinematic dipole. For a better comparison we therefore also subtract the lensing dipole even though it is
taken into account in our analysis.}
\end{figure*}

\subsection{Ray Tracing \& Map Making}\label{subsec:raytracing}

In order to
compute the ISW temperature anisotropies from the DEUS-FUR light-cone data we use the ray-tracing algorithm part of the fast parallel C++ AMR library \textit{Magrathea} developed in \cite{Reverdy2014}. For each ray a past-null direction is chosen at the observer, and the geodesic equations, consisting of
a coupled set of ordinary differential equations, are numerically integrated backwards in time. Following the true photon path, the gravitational potential and its gradient are obtained from the adaptive mesh by multilinear (cloud-in-cell) interpolation. The set of differential equations is then solved by a Runge-Kutta fourth-order method with adaptive time steps (four time steps per AMR cell). Our methodology is very general as long as the time derivative of the potential can be measured from the gravity light cone. In particular, we do not need to specify any relation between the matter density field and the gravitational potential, which means that a wide range of dark energy and modified gravity models can be studied without changing the analysis pipeline. We consider this approach to be an improvement over previous ones also because the light cones cover the full sky without replica, and the geodesic integration is done at the resolution of the AMR simulation.

DEUS-FUR was not designed
to specifically investigate the ISW-RS effect, and because of this the time derivative of the potential was not stored. Nevertheless, it can
be recovered at large scales (i.e.\ larger than the shell size) from the gravitational field (the spatial gradient of the potential computed at a given time) and a total derivative (computed along the light cone). More specifically, we can rewrite the partial time derivative $\partial\psi/\partial\tau$ as
\begin{equation}
 \frac{\partial \psi}{\partial \tau} = \frac{d\psi}{d\tau} - \frac{\partial\psi}{\partial x^i} n^i\,,
\end{equation}
from which we have that
\begin{equation}
 -2\int\limits_\mathrm{src}^\mathrm{obs} \frac{\partial \psi}{\partial \tau} d\chi = -2 \psi_\mathrm{obs} + 2 \psi_\mathrm{src} + 2 \int\limits_\mathrm{src}^\mathrm{obs} \frac{\partial \psi}{\partial x^i} n^i d\chi\,,
\end{equation}
where we have computed the spatial gradient of the potential on the mesh assuming a five-point stencil finite difference approximation.

We have generated full-sky maps of the ISW temperature anisotropies of resolution $N_\mathrm{side}$ by numerically integrating the geodesic equation for all the $12 \times N_\mathrm{side}^2$ light rays received from the directions of the HEALPix pixels \cite{Gorski:2004by}. Here, we specifically set $N_\mathrm{side}$ = $512$. In addition to
the ISW signal we also compute
the lensing deflection angle, which allows us to reconstruct the CMB lensing potential that we will discuss in Section \ref{subsec:lensing}. To this purpose we first generate a map of the deflection vector, then we extract the lensing potential as the generator of its curl-free part. A small curl part is also present due to higher-order lensing, but we do not study this here.

In Fig.\ \ref{fig:maps} we show full-sky maps of the ISW temperature anisotropies (left panels) and the lensing potential (right panels) generated from the light cones of the three DEUS-FUR cosmological models. In the next Section we use the HEALPix package \cite{Gorski:2004by} to evaluate angular correlation functions of the full-sky maps.

\section{Results}\label{sec:results}

\subsection{ISW-Temperature Power Spectrum}

\begin{figure*}[tb]
\begin{minipage}{\textwidth}
 \centerline{\includegraphics[width=0.85\columnwidth]{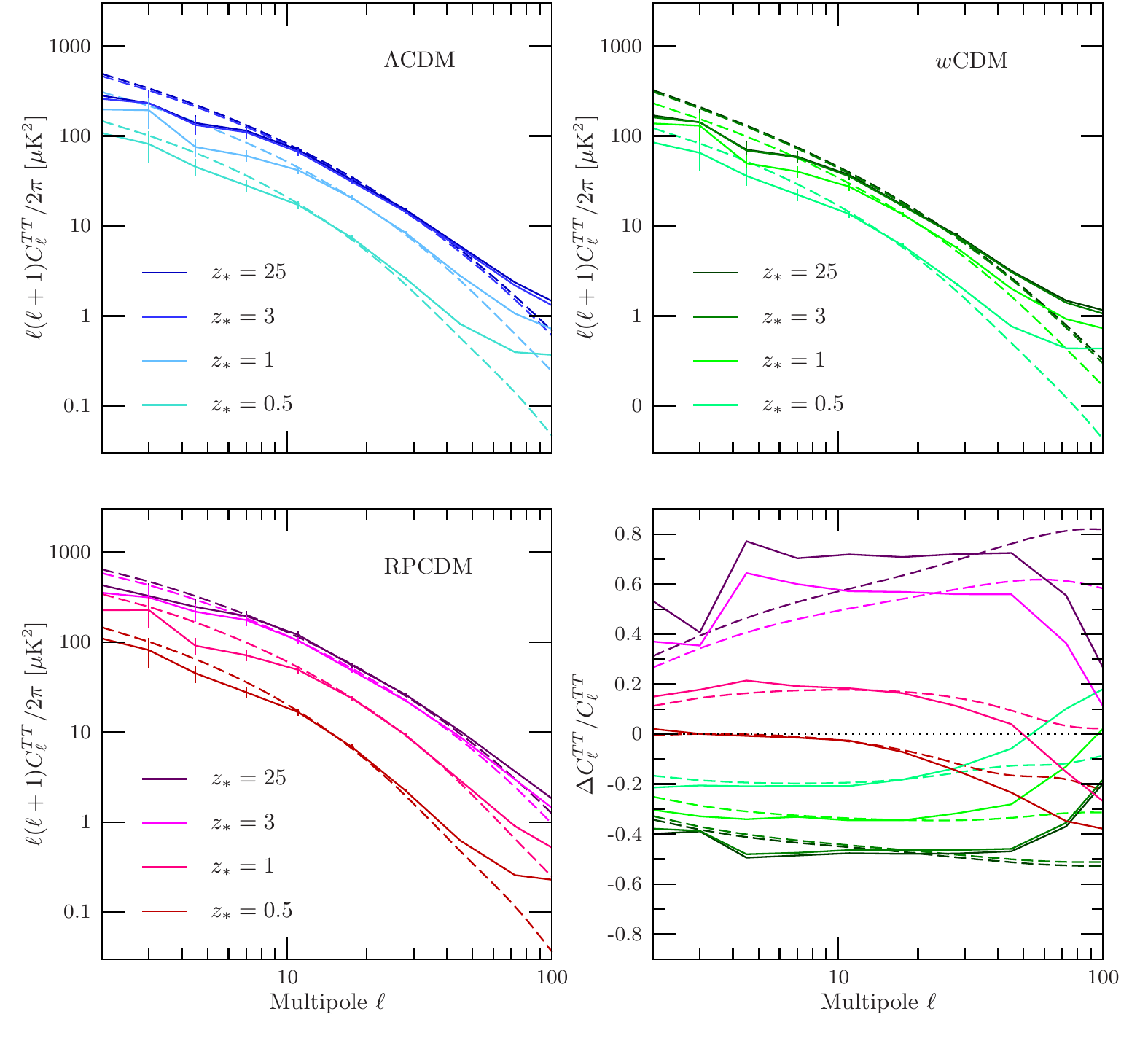}}
\end{minipage}
\caption{\label{fig:isw} \small Angular power spectra of the ISW signal for three different cosmologies, using logarithmic $\ell$-band powers. In order to show where the signal is mostly generated, the temperature anisotropy is calculated by integrating along the line of sight from the observer (at $z=0$) to various finite distances $z_\ast$. For the two evolving DE models ($w$CDM and RPCDM) the lower right panel shows the relative difference with respect to $\Lambda$CDM. In all panels, dashed lines indicate the predictions from linear theory as computed by CLASS.
}
\end{figure*}

In Fig.~\ref{fig:isw} we plot the angular power spectrum obtained from the estimator
\begin{equation}
 C_\ell^{TT} = \frac{1}{2\ell + 1} \sum\limits_m \vert a^{TT}_{\ell m} \vert^2\,,
\end{equation}
where $a^{TT}_{\ell m}$ are the coefficients of the spherical harmonic decomposition of the full-sky maps of the temperature anisotropies as computed using the 
 the HEALPix package. For each cosmology, we compute the late ISW-RS signal for four different starting redshifts $z_\ast$, integrating the photon trajectories between $z_\ast$ and $z=0$. The error bars show statistical errors after combining the estimators from the light cones of two independent observers and binning the data logarithmically in $\ell$.
From the dependence of the signal on $z_*$, we can see that the differences among the various model predictions correlate with the differences in the redshift evolution of the linear growth rate shown in Fig.~\ref{fig:cosmo}. More specifically, we can see that in the $w$CDM case (upper right panel) the bulk of the signal is generated at low redshift. As an example, between $z_\ast=1$ and $3$ the ISW signal at $\ell \sim 10$ only increases by $16\%$, while in the same redshift interval the signal increases by $20\%$ in the $\Lambda$CDM case and $40\%$ for RPCDM. This is consistent with the fact that at these redshifts the linear growth rate of the $w$CDM model remains closer to the Einstein--de~Sitter value ($f \sim 1$), than $\Lambda$CDM and RPCDM, respectively. Moreover, as shown in Fig.~\ref{fig:cosmo}, the linear growth rate of $w$CDM (RPCDM) is systematically higher (lower) than the $\Lambda$CDM case, thus causing a smaller (larger) ISW signal relative to the $\Lambda$CDM prediction as shown in the lower right panel of Fig.~\ref{fig:isw}.

Overall, we find a good agreement with the linear theory for multipoles $\ell \lesssim 50$ where the linear ISW effect is expected to dominate over the non-linear RS effect. Instead over the range $60\lesssim \ell \lesssim 100$ we notice a change of the slope of the power spectrum and a departure from the linear theory that is cosmological model dependent. This departure marks the onset of the non-linear RS effect. It is worth remarking that while the amplitude of the ISW power spectrum is mostly determined by the linear growth rate through the ($1-f$)-factor in Eq.~(\ref{eq:limber}), the scale of deviation from the linear theory primarily depends on the amplitude of the matter density power spectrum (i.e. $\sigma_8 D_+$, where $D_+$ is the linear growth factor). We can see that at multipoles $\ell \sim 100$ the trends flatten to a plateau. However, at these multipoles our estimation of the temperature fluctuations is dominated by noise that results from the limitations of the numerical computation. In fact, our integration method evaluates the signal from discretized data. In particular, as explained earlier, $\partial\psi/\partial\tau$ is not directly computed during the simulation run, but estimated from the total derivative and the spatial gradient of the potential over the mesh. The signal is
therefore effectively recovered by summing the jumps of the gradient between the time steps of the simulation, or in other words by sampling from a relatively small number of locations. This procedure prevents an exact evaluation of the temperature fluctuations on scales smaller than the sampling rate. Consequently, at the multipoles corresponding to these scales the correlation of the numerical noise with itself (whose spectral properties are consistent with a Poisson process) provides the dominant contribution to the power spectrum,  leaving the signal of the RS effect embedded in noise.

\subsection{ISW-Matter Density Correlation}

\begin{figure*}[tb]
\begin{minipage}{\textwidth}
 \centerline{\includegraphics[width=0.85\columnwidth]{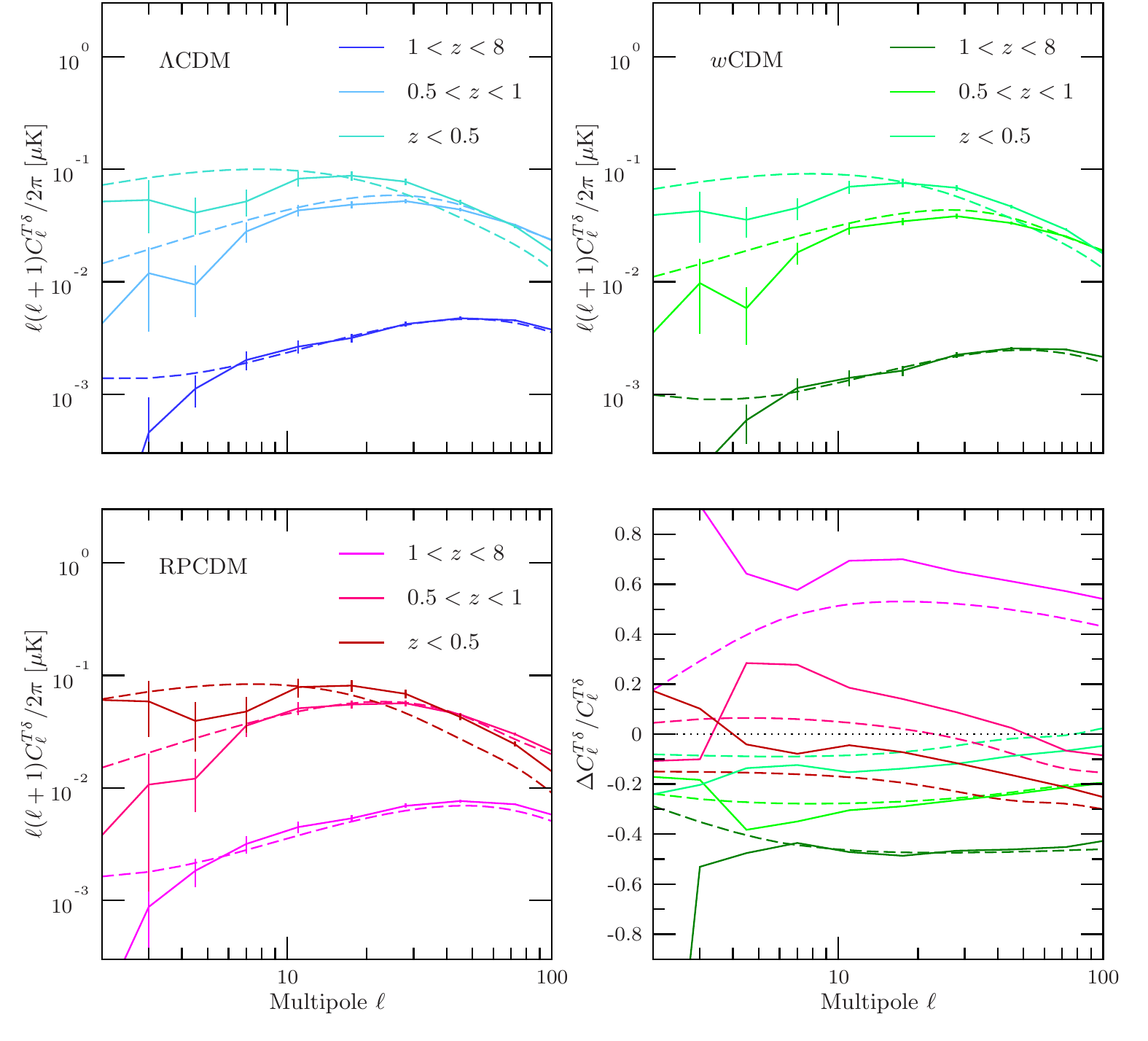}}
\end{minipage}
\caption{\label{fig:iswXmatter} \small Angular cross power spectra of the ISW signal with the matter density contrast for three different redshift bins, using logarithmic $\ell$-band powers. For the two evolving DE models ($w$CDM and RPCDM) the lower right panel shows the relative difference with respect to $\Lambda$CDM. Dashed lines indicate the predictions from linear theory as computed by CLASS.}
\end{figure*}

\begin{figure*}[tb]
\includegraphics[scale=1]{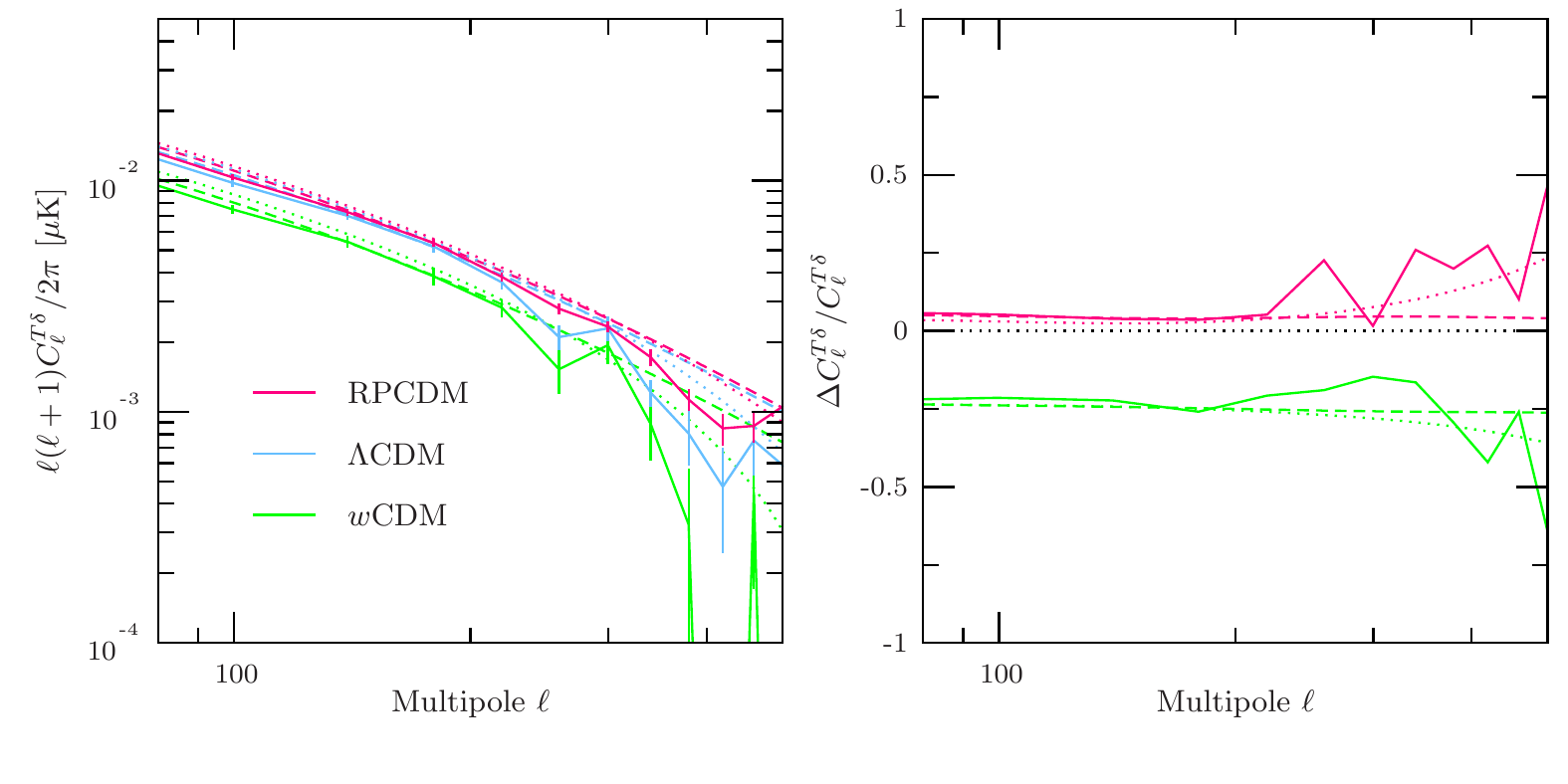} 
\caption{\label{fig:nonlinear} \small The left panel shows the angular cross-power spectra between CMB temperature and matter density contrast (integrated between $z=0$ and $z=9$) at small scales or large multipoles ($\ell > 80$) where nonlinear effects become important. The right panel shows the relative difference of the cross-power spectra to the $\Lambda$CDM linear prediction.
The different lines correspond to the DEUS-FUR numerical simulation (solid lines),  linear prediction (dashed lines) and  HALOFIT prediction (dotted lines).
}
\end{figure*}

As already mentioned, the ISW signal in the CMB temperature anisotropy auto-power spectrum cannot be disentangled from the contribution of other processes which generate temperature fluctuations at the last scattering surface. Instead, it can be detected by cross-correlating CMB temperature maps with the late-time distribution of cosmic structures. Temperature anisotropies generated at early time are largely uncorrelated with the spatial distribution of matter density perturbations at late time. In contrast, ISW anisotropies are sourced by the time varying gravitational potentials associated to late-time matter density perturbations which are traced by cosmic structures.

Here, we compute the cross-correlation between the ISW signal and the matter distribution by constructing full-sky maps of the average density contrast in a given distance interval of the observer's light cone, namely:
\begin{equation}
 \delta(\mathbf{n}) = \frac{\int W(\chi) \delta(\mathbf{n} \chi) d\chi}{\int W(\chi) d\chi}\,,
\end{equation}
where $W(\chi)$ is a tophat selection function that specifies the distance interval. Despite the fact that we use the information recorded along the past light cone, it
should be noted that the density contrast we are evaluating is not a directly observable quantity. First, we use the distribution of dark matter particles to define $\delta$, while observations of large-scale structure have to rely on biased tracers such as galaxies. Second, for simplicity we do not take into account relativistic projection effects such as
e.g.\ weak lensing that would perturb the observed number density of tracers in a given solid angle element. While it may seem that this could be rectified by making use of our
ray tracing algorithm, we remind the reader that the particle positions are not provided in the appropriate gauge\footnote{As our ray tracer uses
longitudinal gauge it requires the number count per \textit{coordinate} volume in that gauge. This quantity is different in different gauges and only relates to a gauge-invariant number count per \textit{redshift space} volume after the relativistic projection effects have been included.} for this task. This gauge issue is not very important on small scales, but it could contaminate the large angular scales we are mostly interested in \cite{Adamek:2019aad}. The quantity $\delta(\mathbf{n})$ should therefore be understood merely as an ideal theoretical probe that is easy to compute and that can be used as a proxy for assessing
the ISW-matter cross-correlation in different dark energy models.

Fig.~\ref{fig:iswXmatter} shows the ISW-matter cross-correlation power spectrum of the three DEUS-FUR models for three redshift bins covering the range $0\leq z<0.5$, $0.5\leq z<1$, and $1\leq z<8$ respectively. We use a power spectrum estimator similar to that described in the previous section and again applied to the full-sky maps of two independent observers in each case. We also show the theoretical model predictions obtained from linear theory (dashed lines). As expected, for all three cosmologies the cross-correlation signal is larger at lower redshift, and the relative contribution of the different redshift bins follows the trends already discussed in the previous section. Higher multipoles are systematically more correlated at higher redshift which is simply a geometric effect. The lower-right panel of Fig.~\ref{fig:iswXmatter} shows the relative difference of the cross-spectra with respect to the $\Lambda$CDM case. 

Overall, we can see that at relative low multipoles ($\ell < 100$) or equivalently at large angular scales, the linear theory reproduces reasonably well the numerical cross-spectra from the DEUS-FUR simulations. The agreement is stronger for the higher redshift bins, while we may notice a departure from the linear theory in the lowest redshift bin at $\ell\approx 100$. The exact angular scale of such departure depends on the underlying cosmological model and marks the onset of non-linearities contributing to the RS effect.

To highlight this point in Fig.~\ref{fig:nonlinear} we plot the cross spectra (left panel) and their relative differences with respect to the linear $\Lambda$CDM prediction (right panel) at high multipoles ($\ell > 100$), where we have integrated the signal between $z=0$ and $z=9$ to have a larger signal-to-noise. Notice that as we explore smaller scales, we approach the spatial resolution of the simulations. This manifests in a drop of the cross-power spectra at large multipoles and in the presence of noise. However, these spurious numerical effects are mostly independent of the underlying cosmology and cancels out when plotting relative differences. This allows us to recover the signature of non-linearities at small scales which is expected to depend on the cosmological model \cite{Alimi2010}. We can clearly see this in the right panel of Fig.~\ref{fig:nonlinear}, where the differences of the $w$CDM and RPCDM cross-spectra with respect to the linear $\Lambda$CDM prediction increase at larger multipoles and correlate with their $\sigma_8 D^+$ value. We also plot an analytical prediction of the non-linear regime following \cite{Nishizawa2008}. At these small scales, the Limber approximation is applicable and we use Eq.~(\ref{eq:limber}) with 
\begin{equation}
  q_2(\chi)=\frac{1}{\Delta \chi}
\end{equation}
inside the top-hat region of width $\Delta \chi$ and $q_2(\chi)=0$ outside. $P_{\delta \delta}$ in Eq.~(\ref{eq:limber}) becomes the non-linear power spectrum and $f$ becomes the (scale-dependent) non-linear growth rate $f(k,\chi)=\frac{1}{2} d \ln P / d \ln a$. As shown in \cite{Alimi2010}, the imprint of non-linearities on the matter power spectrum depends on the specificities of the underlying dark energy model. However, these are not captured by the standard HALOFIT prescription \cite{Smith2003}. Instead, we use the prescription from \cite{Takahashi2012}, which accounts for dark energy models with a constant equation of state $w$. This covers the $\Lambda$CDM and $w$CDM cases, while for the RPCDM model, we approximate its time varying equation of state with its mean value across the cosmic history, corresponding to $w=-0.83$. Notice though that the imprints of dark energy on non-linear scales play an important role (at several percent level) only beyond modes $k\gtrsim 1\,h$~Mpc$^{-1}$. As we focus on large scales corresponding to $k< 1\,h$~Mpc$^{-1}$, in this regime the prescription by \cite{Takahashi2012} provides accurate enough predictions of the matter power spectra. Moreover, given the level of fluctuations exhibited by the numerical spectra at these scales, non-linear model predictions accurate at the percent level are not needed in our analysis.
In the left panel of Fig.~\ref{fig:nonlinear} we plot the non-linear predictions of the ISW-density cross-spectra for the $\Lambda$CDM, $w$CDM and RPCDM models against the numerical simulation results. As we can see the cosmological dependence of analytical predictions and simulations are similar. In the right panel of Fig.~\ref{fig:nonlinear} we plot the relative differences with respect to the $\Lambda$CDM case. We can see that the onset of the RS effect depends on the model of dark energy, thus confirming previous claims in the literature that such a scale can be used as a probe of dark energy \cite{Nishizawa2008,Lee2015}. More accurate prescriptions for the computation of the non-linear matter power spectrum of dynamical dark energy models are required if we had extended the analysis deep in the non-linear regime. In such a case prescriptions such as those developed in \cite{Casarini2016,Mead2016} are needed. However, this is beyond the scope of this paper.

\subsection{ISW-Lensing Potential Correlation}
\label{subsec:lensing}

\begin{figure*}[tb]
\begin{minipage}{\textwidth}
 \centerline{\includegraphics[width=0.85\columnwidth]{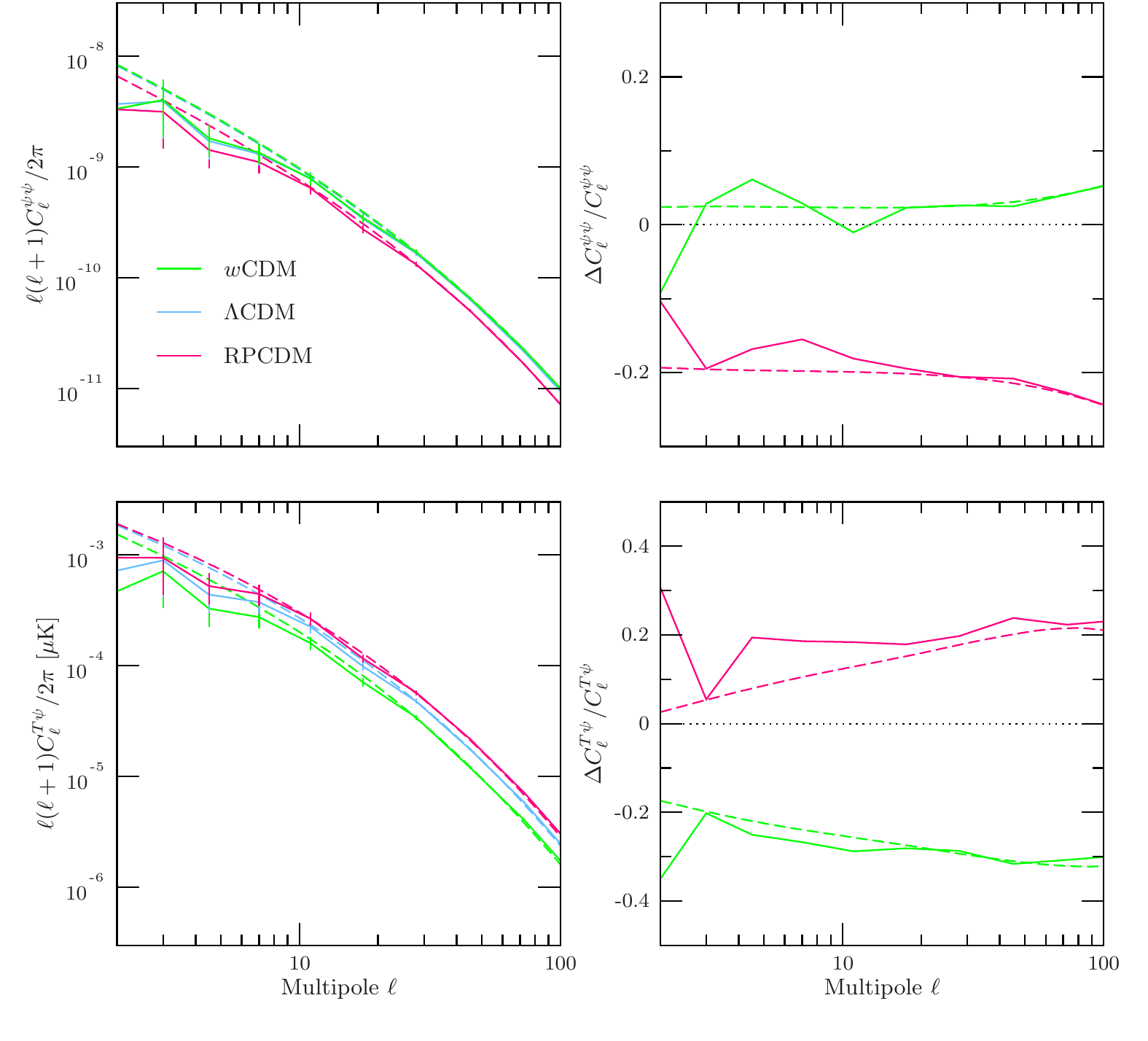}}
\end{minipage}
\caption{\label{fig:lensing} \small Angular power spectra of the lensing potential (top panels) and its cross power spectra with the ISW signal (bottom panels), using logarithmic $\ell$-band powers. Dashed lines are linear predictions computed by CLASS.}
\end{figure*}

The gravitational potentials of large-scale structure
modify the temperature anisotropy pattern through weak gravitational lensing. The lensing potential for the CMB can be constrained from CMB observations directly \cite{PlanckWL}, without the need for external large-scale structure information. Since lensing and ISW effect both originate from the same matter perturbations they are expected to be highly correlated at redshifts where the potentials are decaying.

We construct the lensing potential by taking the curl-free part of the total deflection angle that we compute by integrating the photon geodesic equations for the full sky. In Fig.~\ref{fig:lensing} we plot the angular power spectra of the lensing potentials (top panels) and the cross-correlations with the ISW signal (bottom panels) for our three cosmologies.
The right panels show the relative difference with respect to $\Lambda$CDM. 

Again, for low multipoles ($\ell < 100$) we find an excellent agreement with the predictions from the linear theory (dashed lines) both for the lensing potential auto-correlation function and for the lensing-ISW cross-correlation. This validates again our methodology. An analytical calculation of the lensing-ISW cross-power spectrum following \cite{Nishizawa2008} with Limber approximation (i.e.\ using the lensing weight for $q_2$)  indicates that the scale for non-linearity is beyond $\ell=500$ and therefore at scales below the resolution of our data. This is because the CMB-lensing kernel peaks at high redshift pushing the scale of non-linearities towards larger $\ell$.

The cosmological dependence reaches $30\%$ (between RPCDM and $w$CDM at $\ell=100$) for the auto-correlation of the lensing potential. It appears more pronounced for the cross-correlation between the ISW signal with the lensing potential where it reaches $50\%$ and the order of the spectra is reversed. Moreover while $w$CDM and $\Lambda$CDM are indistinguishable from a lensing perspective, they are more than 1-$\sigma$ away for the ISW-lensing cross-correlation (where $\sigma^2$ is the cosmic variance). This illustrates the complementarity of the lensing and ISW probes.

\begin{figure*}[tb]
\begin{minipage}{\textwidth}
 \centerline{\includegraphics[width=0.85\columnwidth]{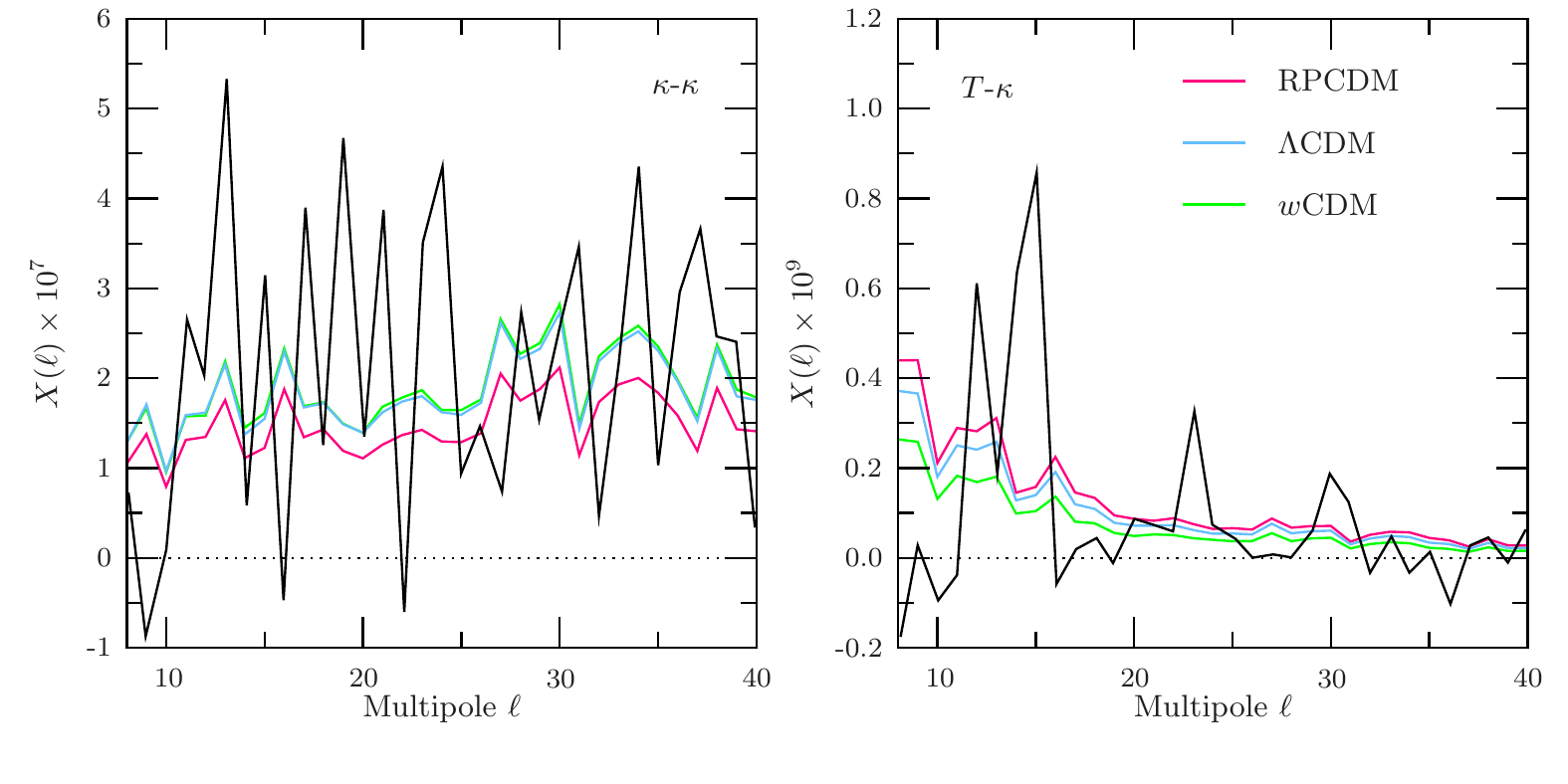}}
\end{minipage}
\caption{\label{fig:compareplanck} \small Comparison of DEUS-FUR cross-spectra $X(l)$ for $\kappa$-$\kappa$ (left) and $T$-$\kappa$ (right) to the Planck measurement (solid black line) from \cite{PlanckISW} (we follow the convention of that paper). The three models agree with the data. In the future, reducing systematics and the use of multiple probes will allow to disentangle the three ``realistic'' cosmologies studied in this paper.}
\end{figure*}

Finally, in Fig.~\ref{fig:compareplanck} we confront our CMB lensing and ISW spectra to measurements from the Planck collaboration \cite{Planck}. Given the observational systematics, the measurement errors are not down to cosmic variance. The three models are therefore compatible with the current measurements. It will however be possible in the future to lower the systematics, combine several probes (ISW-density, ISW-lensing) at several redshift, and explore the RS effects. Altogether this should allow the discrimination between various cosmological models compatible with CMB and supernovae data.

\section{Conclusions}\label{sec:conclusions}

We have presented a study of the ISW-RS effect using data from the Dark Energy Universe Simulation -- Full Universe Runs of three different dark energy models. As these simulations encompass the full
observable volume of the simulated cosmologies, we have been able to perform a thorough analysis of the ISW signal using full-sky light-cone data with a photon propagation across the cosmic non-linear matter distribution up to very
high redshifts and for two observers located at two different space-time positions. The usual spurious effects inherent to replica methods are thus completely avoided. Moreover, we used a sophisticated ray-tracing technique that solves the photon geodesic equation along the photon trajectory without recourse to the Born approximation. Our methodology is very general and applicable to both large and small scales.

The cosmological models of the DEUS-FUR simulations are a flat $\Lambda$CDM model, a quintessence model with Ratra-Peebles potential (RPCDM) and a phantom dark energy model with constant equation of state $w=-1.2$ ($w$CDM). We extracted the angular correlation functions from full-sky maps for such {\it realistic} cosmological models. The auto-correlation functions of ISW signal are very weakly dependent on cosmology for small multipoles ($\ell \ll 50$) and in very good agreement with the linear prediction. For multipoles between $\ell = 50$ and $\ell = 100$, we can see the onset of the RS effect. At larger multipoles the numerical noise dominates the signal. This is caused by the way our integration method extracts the signal from the discretized data for the time derivative of the gravitational potential which was not specifically stored in our simulations. Therefore we can only observe the beginning of the transition to the RS effect as the non-linear signal is then drowned in the numerical noise. 

The ISW-RS signal in the CMB anisotropy power spectrum is not directly observable due to the contribution of early-time processes which generate temperature anisotropies over the same range of angular scales. Nevertheless, it can be detected through cross-correlation of CMB temperature anisotropy maps with the distribution of large-scale structures. This is because the late-time varying gravitational potentials which sources the ISW-RS effect are traced by cosmic structures. Here, we have performed an analysis of the ISW-matter correlation from the DEUS-FUR light-cone data.
We find that the differences among the cross-spectra of the DEUS-FUR models to be enhanced at $\ell \gtrsim 100$.
In particular, we clearly find a deviation from the linear prediction as non-linear effects becomes dominant at small scales. Such effects are expected to be cosmological model dependent and do account for the observed differences between the ISW-correlation spectra of the DEUS-FUR models.
In fact, the linear growth rate of matter density perturbations is larger in $w$CDM than in the $\Lambda$CDM case and even more compared to RPCDM. Consequently the cosmic structure formation is more efficient in $wCDM$ than in $\Lambda$CDM and RPCDM, respectively. This results in an attenuation of the ISW-matter correlation signal in the $w$CDM model with respect to the $\Lambda$CDM and RPCDM cases, mainly due to a lower ISW amplitude, and a decrease on the scale of deviation from the linear regime of matter clustering. 

We have also investigated the correlation of the ISW effect with the lensing potential from the matter distribution. Again, at low multipoles we find the results to be well reproduced by the linear theory. We find the cross-correlation to be particularly sensitive to the underlying cosmological model with the amplitude of the signal varying more than the cosmic variance error among the DEUS-FUR models. The comparison of our results to ISW and lensing data from {\it Planck} shows that these measurements cannot yet disentangle between the three dark energy models studied here.
However, in the light of the observational data that will be accessible with future survey programs, the ISW-lensing correlation in combination with the detections of the non-linear features of the ISW-LSS correlation are likely to provide a powerful cosmological proxy and a probe of the nature of dark energy.

\bigskip
 
\begin{acknowledgments}
We thank V. Reverdy and M.-A. Breton for fruitful discussions and assistance with the use of \textit{Magrathea}.
This work was granted access to the HPC resources of TGCC under the allocation 2017 -- 042287 made by GENCI. The DEUS project is supported by the \textit{Domaine d'Int\'er\^{e}t Majeur en Astrophysique et Conditions d'Apparition de la Vie} (DIM ACAV) of the R\'egion \^{I}le-de-France. The research leading to these results has received partly funding from the European Research Council under the European Community Seventh Framework Program (FP7/2007-2013 Grant Agreement no. 279954) ERC-StG ``EDECS''. JA additionally acknowledges funding by STFC Consolidated Grant ST/P000592/1 during the final stages of this project.
\end{acknowledgments}


\begin{thebibliography}{11}
\expandafter\ifx\csname natexlab\endcsname\relax\def\natexlab#1{#1}\fi
\expandafter\ifx\csname bibnamefont\endcsname\relax
  \def\bibnamefont#1{#1}\fi
\expandafter\ifx\csname bibfnamefont\endcsname\relax
  \def\bibfnamefont#1{#1}\fi
\expandafter\ifx\csname citenamefont\endcsname\relax
  \def\citenamefont#1{#1}\fi
\expandafter\ifx\csname url\endcsname\relax
  \def\url#1{\texttt{#1}}\fi
\expandafter\ifx\csname urlprefix\endcsname\relax\def\urlprefix{URL }\fi
\providecommand{\bibinfo}[2]{#2}
\providecommand{\eprint}[2][]{\url{#2}}

\bibitem{SachsWolfe1967} R.K.\ Sachs and A.M.\ Wolfe, Astrophys.\ J.\ \textbf{147}, 73 (1967).
\bibitem{CrittendenTurok1996} R.G.\ Crittenden and N.\ Turok, Phys.\ Rev.\ Lett.\ \textbf{76}, 575 (1996).
\bibitem{WMAP} C.L.\ Bennett {\it et al.}, Astrophys.\ J.\ Supp.\ S.\ \textbf{148}, 1 (2003); G.\ Hinshaw {\it et al.}, Astrophys.\ J.\ Supp.\ S.\ \textbf{170}, 288 (2007); C.L.\ Bennett \textit{et al.}, Astrophys.\ J.\ Supp.\ S.\ \textbf{208}, 20 (2013).
\bibitem{Planck} Planck Collaboration: P.A.R.\ Ade {\it et al.}, Astron.\ \& Astrophys.\ \textbf{571}, A1 (2014); Planck Collaboration: R.\ Adam {\it et al.}, Astron.\ \& Astrophys.\ \textbf{594}, A1 (2016).
\bibitem{ISWdetections} P.\ Fosalba, E.\ Gaztna\~naga and F.J.\ Castander, Astrophys.\ J.\ \textbf{597}, L89 (2003); C.\ Boughn and R.G.\ Crittenden, Nature \textbf{427}, 45 (2004); M.R.\ Nolta {\it et al.}, Astrophys.\ J.\ \textbf{608}, 10 (2004); P.\ Vielva, E.\ Mart\'{i}nez-Gonz\'{a}lez and M.\ Tucci, Mon.\ Not.\ Roy.\ Astron.\ Soc.\ \textbf{365}, 891 (2006); T.\ Giannantonio {\it et al.}, Phys.\ Rev.\ D\textbf{74}, 063520 (2006); A.\ Rassat, K.\ Land, O.\ Lahav and F.B.\ Abdalla, Mon.\ Not.\ Roy.\ Astron.\ Soc.\ \textbf{377}, 1085 (2007); T.\ Giannantonio, R.G.\ Crittenden, R.\ Nichol and A.J.\ Ross, Mon.\ Not.\ Roy.\ Astron.\ Soc.\ \textbf{426}, 2581 (2012); Planck Collaboration: P.A.R.\ Ade {\it et al.}, Astron.\ \& Astrophys.\ \textbf{571}, A19 (2014).
\bibitem{PlanckISW} Planck Collaboration: P.A.R.\ Ade {\it et al.}, Astron.\ \& Astrophys.\ \textbf{594}, A21 (2016).
\bibitem{ISWanalyses} P.-S.\ Corasaniti, T.\ Giannantonio and A.\ Melchiorri, Phys.\ Rev.\ D\textbf{71}, 123521 (2005); N.\ Padmanabhan {\it et al.}, Phys.\ Rev.\ D\textbf{72}, 043525; J.D.\ McEwen, P.\ Vielva, M.P.\ Hobson, E.\ Marti\'{i}nez-Gonz\'{a}lez and A.N.\ Lasenby, Mon.\ Not.\ Roy.\ Astron.\ Soc.\ \textbf{376}, 1211 (2007); S.\ Ho, C.\ Hirata, N.\ Padmanabhan, U.\ Seljak and N.\ Bahcall, Phys.\ Rev.\ D\textbf{78}, 043519 (2008); J.-Q.\ Xia, M.\ Viel, C.\ Baccigalupi and S.\ Matarrese, J.\ Cosmol.\ Astropart.\ Phys.\ 09\textbf{09}, 003 (2009).
\bibitem{Pogosian2005} L.\ Pogosian, P.-S.\ Corasaniti, C.\ Stephan-Otto, R.\ Crittenden and R.\ Nichol, Phys.\ Rev.\ D\textbf{72}, 103519 (2005).
\bibitem{Douspis2008} M.\ Douspis, P.G.\ Castro, C.\ Caprini and N.\ Aghanim, Astron.\ \& Astrophys.\ \textbf{485}, 395 (2008).
\bibitem{Ballardini2017} M.\ Ballardini, D.\ Paoletti, F.\ Finelli, L.\ Moscardini, B.\ Sartoris and L.\ Valenziano, Mon.\ Not.\ Roy.\ Astron.\ Soc.\ \textbf{482}, 2670 (2019).
\bibitem{Stolzner2018} B.\ St\"{o}lzner, A.\ Cuoco, J.\ Lesgourgues and M.\ Bilicki, Phys.\ Rev.\ D\textbf{97}, 063506 (2018).
\bibitem{ReesSciama1968} M.J.\ Rees and D.W.\ Sciama, Nature \textbf{217}, 511 (1968).
\bibitem{Puchades2006} N.\ Puchades, M.J.\ Fullana, J.V.\ Arnau and D.\ S\'{a}ez, Mon.\ Not.\ Roy.\ Astron.\ Soc.\ \textbf{370}, 1849 (2006).
\bibitem{Cai2009} Y.-C.\ Cai, S.\ Cole, A.\ Jenkins and C.S.\ Frenk, Mon.\ Not.\ Roy.\ Astron.\ Soc.\ \textbf{396}, 772 (2009).
\bibitem{Smith2009} R.E.\ Smith, C.\ Hern\'{a}ndez-Monteagudo and U.\ Seljak, Phys.\ Rev.\ D\textbf{80}, 063528 (2009).
\bibitem{Cai2010} Y.-C.\ Cai, S.\ Cole, A.\ Jenkins and C.S.\ Frenk, Mon.\ Not.\ Roy.\ Astron.\ Soc.\ \textbf{407}, 201 (2010).
\bibitem{Watson2014} W.A.\ Watson {\it et al.}, Mon.\ Not.\ Roy.\ Astron.\ Soc.\ \textbf{438}, 412 (2014).
\bibitem{Carbone2016} C.\ Carbone, M.\ Petkova and K.\ Dolag, J.\ Cosmol.\ Astropart.\ Phys.\ 16\textbf{07}, 034 (2016).
\bibitem{Carbone2008} C.\ Carbone, V.\ Springel, C.\ Baccigalupi, M.\ Bartelmann and S.\ Matarrese, Mon.\ Not.\ Roy.\ Astron.\ Soc.\ \textbf{388}, 1618 (2008).

\bibitem{Alimi2012} J.-M.\ Alimi {\it et al.}, IEEE Computer Soc.\ Press, CA, USA, SC2012, art.\ 73, arXiv e-prints, arXiv:1206.2838 (2012).  
\bibitem{Reverdy2015} V. Reverdy  {\it et al.}, The International Journal of High Performance Computing Applications, \textbf{29}, 249 (2015). 
\bibitem{Fidler2015} C.\ Fidler, C.\ Rampf, T.\ Tram, R.\ Crittenden, K.\ Koyama and D.\ Wands, Phys.\ Rev.\ D\textbf{92}, 123517 (2015).
\bibitem{Adamek:2017kir} J.\ Adamek, Phys.\ Rev.\ D\textbf{97}, 021302 (2018).

\bibitem{Ratra1988} B.\ Ratra and P.J.E.\ Peebles, Phys.\ Rev.\ D\textbf{37}, 3406 (1988).
\bibitem{Caldwell2003} R.R.\ Caldwell, M.\ Kamionkowski, N.N.\ Weinberg, Phys.\ Rev.\ Lett.\ \textbf{91}, 071301 (2003).
\bibitem{Spergel2007} D.N.\ Spergel {\it et al.}, Astrophys.\ J.\ Supp.\ \textbf{170}, 377 (2007).
\bibitem{Kowalski2008} M.\ Kowalski {\it et al.}, Astrophys.\ J.\ \textbf{686}, 749 (2008).
\bibitem{Alimi2010} J.-M.\ Alimi {\it et al.},  Mon.\ Not.\ Roy.\ Astron.\ Soc.\ \textbf{401}, 775 (2010). 
\bibitem{Bouillot2015} V.\ Bouillot {\it et al.}, Mon.\ Not.\ Roy.\ Astron.\ Soc.\ \textbf{450}, 145 (2015).

\bibitem{Lesgourgues2011} J.\ Lesgourgues, arXiv e-prints, arXiv:1104.2932 (2011).

\bibitem{Prunet2008} S.\ Prunet, C.\ Pichon, D.\ Aubert, D.\ Pogosyan, R.\ Teyssier and S.\ Gottloeber, Astrophys.\ J.\ Supp.\ \textbf{178}, 179 (2008).
\bibitem{Teyssier2002} R.\ Teyssier, Astron.\ \& Astrophys.\ \textbf{385}, 337 (2002).
\bibitem{Roy2014} F.\ Roy, V.\ Bouillot and Y.\ Rasera, Astron.\ \& Astrophys.\ \textbf{564}, A13 (2014).
\bibitem{Guillet2011} T.\ Guillet and R.\ Teyssier, J.\ Comp.\ Phys.\ \textbf{230}, 4756 (2011).

\bibitem{Rasera2014} Y.\ Rasera {\it et al.}, Mon.\ Not.\ Roy.\ Astron.\ Soc.\ \textbf{440}, 1420 (2014).
\bibitem{Fosalba2008} P.\ Fosalba, E.\ Gazta{\~n}aga, F.~J.\ Castander \& M. Manera,  Mon.\ Not.\ Roy.\ Astron.\ Soc.\ \textbf{391}, 435 (2008).
\bibitem{Teyssier2009}R.\ Teyssier,  S.\ Pires, S.\ Prunet \textit{et al.}  Astron.\ \& Astrophys.\ \textbf{497}, 335 (2009). 

\bibitem{Reverdy2014} V.\ Reverdy, PhD thesis, Observatoire de Paris (2014).
\bibitem{Gorski:2004by} K.M.\ G\'orski {\it et al.}, Astrophys.\ J.\ \textbf{622}, 759 (2005).

\bibitem{Adamek:2019aad} J.\ Adamek and C.\ Fidler, J.\ Cosmol.\ Astropart.\ Phys.\ 19\textbf{09}, 026 (2019).
\bibitem{Nishizawa2008} A.J.\ Nishizawa, E.\ Komatsu, N.\ Yoshida \textit{et al.}, Astrophys.\ J.\ \textbf{676}, L93 (2008).
\bibitem[Smith et al.(2003)]{Smith2003} Smith, R.~E., Peacock, J.~A., Jenkins, A., \textit{et al.}, Mon.\ Not.\ Roy.\ Astron.\ Soc.\ \textbf{341}, 1311 (2003).
\bibitem{Takahashi2012} R.\ Takahashi, M.\ Sato, T.\ Nishimichi, A.\ Taruya \& M.\ Oguri, Astrophys.\ J.\ \textbf{761}, 152 (2012).
\bibitem{Lee2015} S.\ Lee, arXiv e-prints, arXiv:1510.04770 (2015).
\bibitem{Casarini2016} L.Casarini, S.A. Bonometto, E. Tessarotto, P.S. Corasaniti, J.\ Cosmol.\ Astropart.\ Phys.\ 16\textbf{08}, 008 (2016).
\bibitem{Mead2016} A.\ Mead, C.\ Heymans, L.\ Lombriser, J.\ Peacock, O.\ Steele \& H.\ Winther,  Mon.\ Not.\ Roy.\ Astron.\ Soc.\ \textbf{459}, 1468 (2016).
\bibitem{PlanckWL} Planck Collaboration: P.A.R.\ Ade {\it et al.}, Astron.\ \& Astrophys.\ \textbf{594}, A15 (2016).


\end{thebibliography}
\end{document}